# LINEture: novel signature cryptosystem

Gennady Khalimov[1][0000-0002-2054-9186] and Yevgen Kotukh[2][0000-0003-4997-620X]

[1] Kharkiv National University of Radio electronics, Kharkiv, Ukraine
[2] Yevhenii Bereznyak Military Academy, Kyiv Ukraine
`yevgenkotukh@gmail.com`

**Abstract.** We propose a novel digital signature cryptosystem that exploits the concept of the brute-force problem. To ensure the security of the cryptosystem, we employed several mechanisms: sharing a common secret for factorable permutations, associating permutations with the message being signed, and confirming knowledge of the shared secret using a zero-knowledge proof. We developed a secret-sharing theory based on homomorphic matrix transformations for factorized permutations. The inverse matrix transformation for computing the shared secret is determined by secret parameters, which results in incompletely defined functionality and gives rise to a brute-force cryptanalysis problem. Randomization of session keys using a message hash and random parameters guarantees the uniqueness of each signature, even for identical messages. We employed a zero-knowledge authentication protocol to confirm knowledge of the shared secret, thereby protecting the verifier against unauthorized signature imposition. The LINEture cryptosystem is built on linear matrix algebra and does not rely on a computationally hard problem. High security is achieved through appropriate selection of matrix transformation dimensions. Matrix computations potentially offer low operational costs for signature generation and verification.

## 1    Introduction

The primary goal formulated in the NIST project is to standardize signatures and KEMs with low costs for signatures, keys, and computation time [1]. Several post-quantum cryptosystems have been proposed [2]. The first four post-quantum cryptographic primitives are Crystals-Dilithium (Dilithium) lattice cryptography [3], SPHINCS+ hash-based cryptography, the McEliece code-based cryptosystem [4], and multivariate public-key cryptography (MPKC) [5].

Lattice-based cryptography security is achieved through the use of NP-hard problems such as the shortest vector problem (SVP), the closest vector problem (CVP), and the shortest independent vector problem (SIVP), as well as learning with errors (LWE) and learning with rounding (LWR) [6–9]. Dilithium relies on the Fiat–Shamir with Aborts framework, as well as SVP [10], to ensure security. The diverse applications of Dilithium require the transmission of both the public key and the signature, prompting a focus on reducing their combined size. Despite abandoning Gaussian sampling, Dilithium achieves the smallest combined signature and public key size among existing lattice-based signature schemes. Classical attacks on signature systems use pairs of incorrect and correct signatures to derive the private key; however, these attacks are limited to deterministic signatures. To protect against such threats, Dilithium employs a randomized approach, guaranteeing the uniqueness of each signature, even for identical messages. Dilithium, recognized by NIST for post-quantum standardization, faces a signature correction attack described in [11], which exploits Dilithium's mathematical structure, along with incorrect signatures and the public key, to derive bits of the secret key. Researchers continue to propose new signature schemes that aim to outperform Dilithium. NIST-selected Dilithium PQC candidates are considered secure and efficient schemes.

The security of hash-based cryptography depends on the collision resistance of cryptographic hash functions. Principal signature schemes include the Merkle signature scheme [12], SPHINCS [13], and the NIST-standardized SPHINCS+ scheme [14]. The underlying assumptions are perceived as considerably more conservative than the structured assumptions underlying Dilithium and Falcon. SPHINCS+ exhibits significantly worse performance than other standards: signature size, verification time, and signing time are one, two, and three orders of magnitude higher, respectively, than those of Dilithium.

Code-based cryptography relies on the NP-hard problem of decoding general linear codes for security, rendering it resistant to both classical and quantum attacks [15]. The principal advantages of code-based cryptographic schemes include well-analyzed complexity assumptions, fast matrix-vector computations, and robust cryptanalytic resilience. However, these schemes face challenges such as large public key sizes (often in the hundreds of kilobytes), increased ciphertext size, and, in some variants, a non-zero decryption failure rate. Classic McEliece requires the highest computational overhead and communication cost due to its large public key size, while producing the smallest ciphertext. Classic McEliece is the slowest scheme for key generation.

Multivariate public-key cryptography (MPKC) is a category of cryptographic schemes based on the difficulty of solving systems of multivariate polynomial equations over finite fields. MPKC possesses several attractive properties: it provides high theoretical resistance to quantum attacks, enables high-speed signing and verification processes due to the use of simple algebraic operations over small fields, and allows for compact signature sizes (for example, Unbalanced Oil and Vinegar (UOV) signatures are typically around 100 bytes). Disadvantages include significantly large public key sizes—often reaching hundreds of kilobytes—vulnerability to algebraic and structural attacks [16], and the lack of a robust and efficient encryption scheme. Despite these shortcomings, MPKC remains an active area of post-quantum cryptography research.

The high overhead of post-quantum algorithms stems from the fact that achieving cryptographic security requires a significant expansion of the signature and public key space. For cryptosystems based on NP-hard problems, this is an inevitable consequence. Cryptosystems of this type lack formal proof of resistance to quantum cryptanalysis, and this may be considered a persistent threat.

To address the problem of constructing a post-quantum cryptosystem with low implementation costs while meeting NIST security requirements, we propose constructing a signature cryptosystem using a novel concept based on brute-force problems with equiprobable solutions. Brute-force problems arise when it is computationally easy to construct public keys from secret keys, while the inverse computations are underdetermined due to the cryptosystem's secret parameters. Classical secret-sharing schemes are algorithms of this type. It is impossible to construct a shared secret without knowledge of the secret parameter from which private shares are constructed. An example is the well-known Shamir threshold scheme, which is based on polynomial interpolation from point values. The security of Shamir's scheme is guaranteed by the properties of polynomials, and an attack on the shared secret is possible only through brute-force methods.

## 2      Our contribution

We develop a theory for constructing asymmetric cryptosystems for which security is determined by a brute-force problem. To ensure the security of the cryptosystem, we employed several mechanisms: sharing a common secret for factorable permutations, associating permutations with the message being signed, and confirming knowledge of the shared secret using a zero-knowledge proof. We developed a secret-sharing theory based on homomorphic matrix transformations for factorable permutations. The inverse matrix transformation for computing the shared secret is incompletely defined, which constitutes one of the conditions of the brute-force cryptanalysis problem. Using the message as one of the parameters in constructing the shared secret represents another condition of the brute-force problem, since the inverse permutations for each signature are unique, thereby removing the restriction to deterministic signatures. Randomization of session keys guarantees the uniqueness of each signature, even for identical messages. We applied a zero-knowledge authentication protocol within the Fiat–Shamir paradigm to prove knowledge of the shared secret, thereby protecting the verifier against unauthorized signature imposition. Organization. In Section 2, we present a description of the LINEture signature cryptosystem based on secret sharing and identity proof. In Section 3, we present a parameterization of secret substitutions on the elementary abelian 2-group of order $2^m$. In Section 4, we describe the construction of private shares and the shared secret based on homomorphic matrix transformations. In Section 5, we perform security analysis and provide complexity estimates for basic brute-force and analytical attacks. Section 6 provides security estimates and implementation costs.

## 3      Introducing LINEture cryptosystem

The LINEture signature cryptosystem is built on a mechanism for secret sharing using substitution vectors for m-bit words. The permutations serve as private shares and defined by random matrices with row vectors in the basis of the space representation over the field. The shared secret is also a permutation matrix and is computed via a homomorphic transformation involving matrix multiplication of the private shares by the session keys. To verify a signature, the session key and, accordingly, the shared secret are presented. The shared secret is unique in each session, even for repeated messages. To protect against the imposition of unauthorized signatures, a zero-knowledge proof is employed to confirm knowledge of the private shares. The digital signature cryptosystem is based on linear matrix algebra. Designing the cryptosystem involves the following steps.

### 3.1  A Secret sharing based on homomorphic matrix transformations for factorized permutations

To sign the message msg, we calculate the hash $h=H(msg)$ and represent it as a vector $y = y[l]$ dimensions $l$ with $m$ bit words. Let's define the vector $S = S[l] = |S_1, S_2, ..., S_l|$ - with substitutions $S_i$, $i = \overline{1,l}$ for $m$ bit words.

We define the signature as the calculated inverse substitutions $S^{-1}$ for the vector $y = y[l]$

$$x = S^{-1}(y), \qquad (1)$$

where are the components of the vector $x[l]=[x_1, x_2, ..., x_l]$ represented by

$$x_i = S_i^{-1}(y_i), \quad i=\overline{1,l}. \qquad (2)$$

Substitutions $S_i$, $i=\overline{1,l}$ define by matrices dimensions $[2m \times m]$ whose rows are basis vectors of the space of dimension $m$ over the field $F_2$ elementary Abelian 2-group of order $2 m$. In this construction, permutations are calculated as the sum of the rows of the matrix $S_i[2m \times m]$.

Let the vector $\beta=[\beta_1, \beta_2, ..., \beta_l]$ consist of components that are the concatenation of matrices $\beta_{ij}$ $j=\overline{1,q}$ dimensions $[2m \times m]$

$$\beta_i[2m \times mq] = \beta_{i1} \| \beta_{i2} \| ... \| \beta_{iq}, \quad i=\overline{1,l}. \qquad (3)$$

where $\beta_{ij}[2m \times m]$ are binary matrices, whose rows are vectors of the dimension space $m$ over the field $F_2$

Let $\varpi_j[m \times m]$, $j=\overline{1,q}$ be binary matrices. Concatenate $\varpi_j[m \times m]$ row by row

$$E[m \times mq] = \varpi_1 \| \varpi_2 \| ... \| \varpi_q. \qquad (4)$$

each permutation $S_i$ of $i=\overline{1,l}$ a vector $S$ by a matrix product with bitwise computation over $F_2$

$$S_i[2m \times m] = \beta_i E^T, \quad i=\overline{1,l}. \qquad (5)$$

Matrix multiplication (5) can be represented by a functional expression

$$S_i = \sum_{j=1}^{q} \beta_{ij}\varpi_j , \quad i=\overline{1,l} . \tag{6}$$

Let's define a secret binary matrix $\omega[mq \times mq]$. Let's calculate the permutation vector $B=[B_1,B_2,...,B_l]$

$$B_i[2m \times mq] = \beta_i \omega = |\beta_{i1}\|\beta_{i2}\|...\|\beta_{iq}|\omega , \quad i=\overline{1,l} . \tag{7}$$

Since the matrix $\omega$ is a secret, knowledge of the vector components $B$ does not allow us to find $\beta$.

To restore the secret vector, $S$ we calculate the components

$$S_i = B_i \omega^{-1} E^T = B_i \psi = \beta_i E^T , \quad i=\overline{1,l} \tag{8}$$

where is the matrix equals to

$$\psi[mq \times m] = \omega^{-1} E^T . \tag{9}$$

B substitution $S$ vector is a shared secret whose components $S_i$ are constructed by combining partial secrets $B_i$ by session key $\psi$.

We bind the matrix $E$ with a hash of a random value $r$ and a message $H(r,msg)$. Then, for each signing iteration, the session key $\psi$ will be different and this solution ensures randomization of the secret $S$ for each signature.

In the described cryptosystem, $\omega$ defines as a long-term secret key and the vector $E$ defined as a secret session key. $B$ and $\psi$ defined as public keys respectively.

The general signature scheme is as follows.

The signer calculates a hash for a random value $r$, $H(r,msg)$, constructs the matrix $E$ (4) and calculates the secret $S$ (5). For the vector $y=H(msg)$ given the inverse substitution vector $S^{-1}$, computes the vector $x$, which together with $\psi$ is the signature.

The verifier, given a vector of values, $x$ computes the vector in reverse order $y=B(x)$. It multiplies it $y$ by the session key matrix $\psi$. If $y\psi=H(msg)$, then verification is successful.

### 3.2  Proof and Identity

The $i=\overline{1,l}$ substitutions $S_i$ are bijective. This fact allows us to construct a signature $x$ for $y=H(msg)$. The bijective $S_i$, $i=\overline{1,l}$ can be constructed for random session keys $\psi$ with a small number of attempts. To prove identity, the signer must show that the shared secret is the substitution vector $S$ which has been constructed using the secret keys $\omega$, $E$. Substitutions $S_i$, $i=\overline{1,l}$ are calculated through the matrix product $E$ and $\beta_i$ in (5). If we apply randomization to one or more matrices $\varpi_j$ in $j=\overline{1,q}$ to the vector $E$, then we can construct several vectors $E$ for which the same secret $S_i$, $i=\overline{1,l}$ will be calculated respectively.

Therefore, to prove identity, several session keys $\psi$ must be presented. These keys will give us a same secret $S$ as a results of calculation (8).

Let us consider an identity proof algorithm for verification using two session keys.

Let's construct a matrix $E(h_1, h_{id}) = \varpi_1 \| \varpi_2 \| ... \| \varpi_q$, where $\varpi_1(h_{id})$ and $\varpi_q(h_1)$ built by hashes $h_{id} = H(r_{id}, msg)$, $h_1 = H(r_1, msg)$. Matrices $\varpi_j$, $j = \overline{2, q-1}$ were constructed using matrices $\varpi_1, \varpi_q$ and the parameters of the substitution vector $\beta$.

Let's calculate the session key $\psi(h_1) = \omega^{-1} E(h_1, h_{id})^T$ (9). For $h_2 = H(r_2, msg)$ we construct the second session key $\psi(h_2) = \omega^{-1} E(h_2, h_{id})^T$. In Section 4 it will be shown how to construct matrices $\varpi_j$, $j = \overline{2, q-1}$ through substitutions $[\beta_1, \beta_2, ..., \beta_l]$. At this step we also will get such $\psi_1$ and $\psi_2$ that yield the same shared secret $S$. We'll reduce the dimension of matrix $\psi$ to $[m(q-1) \times m]$. This allows us to include $\varpi_q(h_1)$ and $\varpi_q(h_2)$ in $\psi(h_1)$ and $\psi(h_2)$ explicitly. This inclusion allows us to verify that the session keys are constructed for the message $msg$. The uniqueness of the secret $S$ for each signature is ensured by the hash $h_{id}$.

To verify identity, we calculate $H(r_1, msg)$ and $H(r_2, msg)$, construct $\varpi_q(h_1)$ and, $\varpi_q(h_2)$ and compare with $\varpi_q(h_1) = \psi_q(h_1)$ and $\varpi_q(h_2) = \psi_q(h_2)$. For the signature $x$, we calculate $u = B(x)$, $y_1 = u\psi(h_1)$ and $y_2 = u\psi(h_2)$. Once $y_1 = y_2$, identity verification is confirmed. The number of session keys can generally be equal to $t \geq 2$.

**Table 1.** Signature algorithm with identity proof for $t=2$ session keys.

| | |
|---|---|
| **General parameters** | $m, l, q, t=2$ |
| **Private signature key** | sk : $\omega, \beta$ |
| - generate | $\omega[mq \times mq]$ |
| - build | $\beta = [\beta_1, \beta_2, ..., \beta_l]$, $\beta_i[2m \times mq] = \beta_{i1} \| \beta_{i2} \| ... \| \beta_{iq}$, $i = \overline{1, l}$. |
| **Public signature key** | pk : $B$ |
| - calculate | $B = \beta\omega = [B_1, B_2, ..., B_l]$, $B_i[2m \times mq] = b_{i1} \| b_{i2} \| ... \| b_{iq}$, $i = \overline{1, l}$ |
| **Signature** $sig = (x, \psi_1, \psi_2, r_1, r_2)$ | |
| - generate | $r_1, r_2, r_{id}$ |
| - calculate | $h_1 = H(r_1, msg)$, $h_2 = H(r_2, msg)$, $h_{id} = H(r_{id}, msg)$ |
| - generate | $\varpi_1(h_{id})$, $\varpi_q(h_1)$, $\varpi_q(h_2)$ |
| - calculate | $\varpi_j(\beta, h_1, h_{id}))$, $\varpi_j(\beta, h_2, h_{id}))$, $j = \overline{2, q-1}$ |
| - to build | $E(h_1, h_{id}) = [\varpi_1(h_{id}) \| \varpi_2 \| \varpi_3 \| ... \| \varpi_{q-1} \| \varpi_q(h_1)]$ |
| - to build | $E(h_2, h_{id}) = [\varpi_1(h_{id}) \| \varpi_2 \| \varpi_3 \| ... \| \varpi_{q-1} \| \varpi_q(h_2)]$ |
| - calculate | $\psi_1 = \omega^{-1} E(h_1, h_{id})^T$ |

$$\psi_2 = \omega^{-1} E(h_2, h_{id})^T$$
$$S = \beta E(h_1, h_{id})^T$$
$$y = H(msg)$$
$$x = S^{-1}(y)$$

**Signature verification**
**Input:** $msg$, $sig = (x, \psi_1, \psi_2, r_1, r_2)$
- calculate $\quad h_1 = H(r_1, msg)$, $h_2 = H(r_2, msg)$
- to build $\quad \varpi_q(h_1), \varpi_q(h_2)$
- check $\quad \varpi_q(h_1) = \psi_q(h_1)$
- $\quad\quad\quad \varpi_q(h_2) = \psi_q(h_2)$
- calculate $\quad u = B(x)$
$$y_1 = u\psi_1$$
$$y_2 = u\psi_2$$
- check $y_1 = y_2 = H(msg)$

---

To verify identity, we used two session keys $\psi_1$, $\psi_2$, each of which leads to the construction of a shared secret $S$. The construction of submatrices $\varpi_j$ in $E$ is determined by the parameterization of the permutations $\beta_i = \beta_{i1} \| \beta_{i2} \| ... \| \beta_{iq}$ and $S = \beta E^T$. As will be shown in Section 4, the shared secret $S$ is unique across hash sets $h_1$, $h_{id}$ and $h_2$, $h_{id}$. The entropy of session keys is determined by the size of the submatrices $\varpi_j$ in $E$, is equal to $m^2(q-1)$ and easily scales by parameters $m$, $q$ and for $t > 2$ the number of session keys.

### 3.3  Parameterization of substitutions $\beta_i$

Let us define the permutations $S_i$, $i = \overline{1, l}$ through matrices dimensions $[2m \times m]$ whose rows are basis vectors of the space of dimension $m$ over the field $F_2$ elementary Abelian 2-group of order $2^m$.

Let $\varsigma$ be elements of an Abelian group and be defined by $m$-bit strings. Let $r = r_1, r_2, ..., r_m$ be the input $m$ bit string. We define the bits $r_j$ of the string $r$ in terms of spinors $r_j = \begin{vmatrix} 1 - r_j \\ r_j \end{vmatrix}$. For bit, 0 we have a spinor $\bar{0} = \begin{vmatrix} 1 \\ 0 \end{vmatrix}$ and bit 1, a spinor $\bar{1} = \begin{vmatrix} 0 \\ 1 \end{vmatrix}$.

We represent the factorization of an Abelian 2-group of order by $2^m$ by matrix $g$ of bit strings that are pairwise combined into blocks $g = [G_1, G_2, ..., G_m]$

$$G_i = \begin{vmatrix} g(i1)_0, g(i2)_0, \ldots, g(im)_0 \\ g(i1)_1, g(i2)_1, \ldots, g(im)_1 \end{vmatrix}, \quad i=\overline{1,m}. \quad (10)$$

The calculation of the transformation $g$ for $m$-bit word $x$ is easy to define by tensor product

$$g(x) = |\overline{x_1}, \overline{x_2}, \ldots, \overline{x_m}| \otimes |G_1, G_2, \ldots, G_m| = \overline{x_1} \otimes G_1 + \overline{x_2} \otimes G_2 + \ldots + \overline{x_m} \otimes G_m, \quad (11)$$

where

$$\overline{x_j} \otimes G_j = \begin{vmatrix} 1-x_j \\ x_j \end{vmatrix} \otimes \begin{vmatrix} g(j1)_0, \ldots, g(jm)_0 \\ g(j1)_1, \ldots, g(jm)_1 \end{vmatrix} = |g(j1)_0(1-x_j) + g(j1)_1 x_j|, \ldots, |g(jm)_0(1-x_j) + g(jm)_1 x_j|$$

Group factorization by bases determines the structures and block types of the matrix $g$. In the given example, we used blocks of type 2, which defines two basis elements of the finite group in each block. This corresponds to a one-bit element of the row $x$. Blocks with a larger number of basis's elements can be used. If the row $x$ is partitioned into bitwise elements $n$, then the basis blocks must be of type $2^n$, $n<m$. In this case, $\beta(x)$ spinors of size should be used for the calculation in expression (11) $2^n$.

The forward transformation $g(x)=y$ is calculated using the tensor product of the input word and the matrix with the group bases. To compute the inverse transformation, $g^{-1}(y)=x$ one must know the factorization $g$. Secrecy of the group factorization can be ensured through homomorphic transformations of the elements of the basic blocks, merging the basic blocks, permuting them, and permuting the elements within the blocks. The efficiency of such transformations, operational costs, and the secrecy provided are extensively discussed in [17-19].

We construct transformations $g$ based on secret factorization over an Abelian 2-group of order $2^m$, using a set of secret homomorphic transformations [18].

A set of transformations for constructing a secret vector factorization for a simple group $g_1$ with type 2 blocks is as follows.

***Transformation 1 Here are the steps for construction of a secret factorization over an Abelian 2-group of order $2^m$***

- permutation of elements $\rho_1: g_1 \to g_2$ in blocks $G_j$, $j=\overline{1,m}$;

- $\rho_2: g_2 \to g_3$ block rearrangement in array $g_2$;

- adding random bits $\rho_3: g_3 \to g_4$ to block rows $G_j$; $j=\overline{1,m}$

- secret homomorphic transformation based on polynomial multiplication $\rho_4: g_4 \to g_5$, $g_5 = \gamma \cdot g_4$ rows of blocks $G_j$, $j=\overline{1,m}$, where is $\gamma$ a polynomial $\gamma \in F(2^m)$;

- secret homomorphic transformation based on matrix multiplication $\rho_5: g_5 \to g_6$, $g_6 = g_5 \cdot \varphi$ rows of blocks $G_j$, $j=\overline{1,m}$, Where $\varphi$ non-singular binary matrix of dimension $m \times m$;

- secret homomorphic transformation based on matrix addition $\rho_6: g_6 \rightarrow g_7$, $g_7 = g_6 + \tau$ rows of blocks $G_j$, $j=\overline{1,m}$, where is $\tau$ a binary matrix of dimension $m \times m$. Each $j$ row of the matrix $\tau$ is added bitwise to each row of the corresponding block. $G_j$

The result is a transformation $g = [G_1, G_2, ..., G_m]$.

**Example 1.** Let's construct a factorization $g$ with blocks of bases of an Abelian 2-group of type 2. Let $m=6$. Let's define

- a prime factorization of the group $g_1$, which is presented in Table 1.
- permutation matrix $\rho_1 =: [\,110110\,]$ elements in the blocks of the matrix $g_1$,
- permutation matrix $\rho_2 =: [\,340152\,]$ matrix $g_2$ blocks $[G_1, G_2, ..., G_m]$,
- random vectors $\upsilon = [\upsilon_1, \upsilon_2, ..., \upsilon_m]$, $\upsilon_j \in F(2^m)$, $j=\overline{1,m}$ to transform $\rho_3: G_j(i)_4 = G_j(i)_3 + \upsilon_j$, $i=\overline{1,2}$, $j=\overline{1,m}$

$$\upsilon = [\upsilon_1, \upsilon_2, ..., \upsilon_m] = \begin{vmatrix} 101111 \\ 101000 \\ 111001 \\ 010100 \\ 000000 \\ 011110 \end{vmatrix},$$

- a random polynomial $\gamma = 1 + x + x^2 + x^4$ for $\rho_4: G_j(i)_5 = G_j(i)_4 \cdot \gamma$, $i=\overline{1,2}$, $j=\overline{1,m}$
- non-degenerate bit matrix y $\varphi$ for $\rho_5: G_j(i)_6 = G_j(i)_5 \cdot \varphi$, $i=\overline{1,2}$, $j=\overline{1,m}$

$$\varphi_{m \times m} = \begin{vmatrix} 101000 \\ 001010 \\ 110001 \\ 000111 \\ 010000 \\ 111010 \end{vmatrix}$$

- bit matrix y $\tau$ for $\rho_6: g(2i-1)_7 = g(2i-1)_6 + \tau(i)$, $g(2i)_7 = g(2i)_6 + \tau(i)$, $i=\overline{1,m}$, where $\tau(i)$ is $i$ a row of the matrix, $g(2i-1)$, $g(2i)$ is a row of the matrix $g$

$$\tau_{m \times m} = \begin{vmatrix} 1\;1\;100 \\ 1 \\ 0\;1\;1010 \\ 0\;\;10\;\;1 \\ 01 \\ 00\;10\;11 \\ 0100\;1\;0 \\ 1\;0\;1010 \end{vmatrix}$$

The results of the calculations $g$ step by step $\rho_1 \div \rho_6$ are presented in Table 2.

**Table 2.** Results of the transformations performed in Example 1.

| $g=[G_1,...,G_m]$ | $g_1$ | $g_1 \to g_2$ | $g_2 \to g_3$ | $G_j(i)_3 + \upsilon_j$ | $G_j(i)_4 \cdot \gamma$ | $G_j(i)_5 \cdot \varphi$ | $G_j(i)_6 + \tau$ |
|---|---|---|---|---|---|---|---|
| $G_1$ | 000000 | 100,000 | 000000 | 101111 | 001011 | 011011 | 100 01 0 |
|       | 100,000 | 000000 | 001000 | 100111 | 110101 | 011111 | 100 1 10 |
| $G_2$ | 100,000 | 010000 | 000100 | 101100 | 011011 | 010001 | 0 01 0 1 1 |
|       | 010000 | 100,000 | 110000 | 011000 | 100011 | 000010 | 0 11 0 0 0 |
| $G_3$ | 000000 | 000000 | 111000 | 000001 | 101111 | 110100 | 1 0 0 0 0 1 |
|       | 001000 | 001000 | 001001 | 110000 | 100111 | 000101 | 0 1 0 000 |
| $G_4$ | 110000 | 000100 | 100,000 | 110100 | 111000 | 010011 | 01 1 0 00 |
|       | 000100 | 110000 | 000000 | 010100 | 000010 | 010000 | 01 1 0 11 |
| $G_5$ | 100,000 | 010110 | 010000 | 010000 | 011101 | 000110 | 0 1 01 0 0 |
|       | 010110 | 100,000 | 100,000 | 100,000 | 111010 | 000011 | 0 1 00 0 1 |
| $G_6$ | 111000 | 111000 | 010110 | 001000 | 111110 | 000100 | 1 0 1 1 1 0 |
|       | 001001 | 001001 | 100,000 | 111110 | 111001 | 101001 | 0 0 0 0 1 1 |

The computation of the transformation $g(x)=y$ for $m$ a bit word $x$ is determined by the tensor product (11). The transformations $\rho_1 \div \rho_6$ mask the factorization of the group.

The calculation of the inverse transform $\beta^{-1}(y)=x$ is performed through inverse operations $\rho_1^{-1} \div \rho_6^{-1}$ with reduction to a row $z_3$ in a factorizable group $g_3$:

$$y_3 = y\varphi^{-1}\gamma^{-1} + \upsilon_\Sigma, \qquad (12)$$

Where $\upsilon_\Sigma = \sum_{j=1}^{m} \upsilon_j$.

For the string, $y_3$ we apply factorization by a simple group $\beta_1$

$$\beta^{-1}(y_3) = (x_1, x_2, ..., x_m)_3. \qquad (13)$$

Let's get the original data string $x$ after inverse permutations $\rho_1, \rho_2$

$$\rho_1^{-1}\left(\rho_2^{-1}(x_1, x_2, ..., x_m)_3\right) = x_1, x_2, ..., x_m. \qquad (14)$$

Substitutions at a length $m$ of bits have potentially good secrecy characteristics since their number has an estimate of $2^m!$.

The entropy estimate of the number of permutations based on group factorization is determined by randomizing transformations $\rho_1 \div \rho_6$ and is large even for small values $m$ [19-31]. For example, one can restrict oneself to the number of non-singular binary matrices $\varphi$ in $\rho_5$, which has the estimate

$$N = (2^m - 1)(2^m - 2)(2^m - 2^2) \cdots (2^m - 2^{m-1}) \approx 2^{m^2 - 2}.$$

The memory cost of group factorization-based substitutions is equal $2m$ to basis vectors, which is significantly less than that of table-based substitutions.

## 4 Construction of a secret based on homomorphic matrix transformations

The private keys in the cryptosystem are defined by a matrix $\omega$ and vectors of substitutions $S$ for digital signature and for identity proof.

We construct $\omega$ as follows.

***Transformation 2. Here is a steps to construct a matrix $\omega[mq \times mq]$***

- we define: $O[m(q-1) \times m]$ - as a zero matrix and $e[m \times m]$ as an identity matrix;

- we generate $\omega_1[m(q-1) \times m(q-1)]$, a random non-singular binary matrix;

- we'll build $\omega[mq \times mq] = \begin{vmatrix} \omega_1 & O \\ O^T & e \end{vmatrix}$;

The vector $S$ to be defined for each component of the vector $\beta=[\beta_1, \beta_2,...,\beta_l]$ by expression (6) through the sum of the permutation matrices $\beta_{ij}$. The permutation matrices $\beta_{ij}$ are input parameters for computing secret session keys $E(h_1, h_{id})$ and $E(h_2, h_{id})$. Construction of permutation matrices $\beta_{ij}$ to be defined by the following transformations.

***Transformation 3. Here is the steps to construct $\beta=[\beta_1, \beta_2,...,\beta_l]$***

- generate $\delta_j[m \times m]$, $\lambda_j[m \times m]$, $R_{ij}[2m \times m]$, $j=\overline{2,q-1}$, $i=\overline{1,l}$ and $\tau=[\tau_1,\tau_2,...,\tau_l]$, $\tau_i[m \times m]$, $i=\overline{1,l}$ which are random binary matrices;

and $\gamma_j[m \times m]$, $j=\overline{2,q-1}$ which are non-singular random binary matrices

- we build $g_{i1} \xrightarrow{\rho_1 \div \rho_5} R_{i1}$, $g_{i1}$ - a simple group of basis vectors

- we calculate $\beta_{i1}=R_{i1}+\tau_i+\sum_{j=2}^{q-1} R_{ij}\delta_j$, $\beta_{ij}=R_{ij} \times \gamma_j$, $i=\overline{1,l}$, $j=\overline{2,q-1}$, (15)

- we calculate $\beta_{iq}=\sum_{j=2}^{q-1} R_{ij}\lambda_j$ (16)

Substitutions $\beta$ are parameterized by matrices $\delta_j$, $\lambda_j$, $\gamma_j$. These parameters are the same for all components $\beta_i$ of the vector $\beta$.

We define the construction of secret session keys $E(h_1, h_{id})$ and $E(h_2, h_{id})$ by the following transformations.

***Transformation 4. Here is the steps to construct $E(h_1, h_{id})$ and $E(h_2, h_{id})$***

We have $h_1=H(r_1,msg)$, $h_2=H(r_2,msg)$, $h_{id}=H(r_{id},msg)$ as an input data.

- we generate $\varpi_1(h_{id})$, $\varpi_q(h_1)$

- we compute $\varpi_j=(\gamma_j \varpi_q(h_1)+\delta_j \varpi_1(h_{id}))\lambda_j^{-1}$, $j=\overline{2,q-1}$ (17)

- we build $E(h_1,h_{id})=\varpi_1 \| \varpi_2 \| ... \| \varpi_q$ (18)

- we generate $\varpi_q(h_2)$

- we compute $\varpi_j = (\gamma_j \varpi_q(h_2) + \delta_j \varpi_1(h_{id}))\lambda_j^{-1}$, $j=\overline{2,q-1}$

- we build $E(h_2, h_{id}) = \varpi_1 \| \varpi_2 \| ... \| \varpi_q$  (19)

A shared secret $S = \beta E^T$ is determined session keys $E(h_1, h_{id})$ and $E(h_2, h_{id})$. The components of the vector $S$ are calculated using formula (6).

Let's calculate $S_i$ with substitution $\beta_{ij}$ (12), (13) and $E(h_1, h_{id})$  (20)

$$S_1(h) = \beta_i E(h_1, h_{id})^T = \beta_{i1}\varpi_1 + \beta_{i2}\varpi_2 + ... + \beta_{iq}\varpi_q =$$
$$\left| (R_{i1} + \tau_{i1} + \sum_{j=2}^{q-1} R_{ij}\delta_j)\varpi_1 + R_{i2} \times \gamma_2 \varpi_2 + ... + R_{i(q-1)} \times \gamma_{(q-1)}\varpi_{(q-1)} \right| + (\sum_{j=2}^{q-1} R_{ij}\lambda_j \varpi_q = \quad (21)$$

$(R_{i1} + \tau_{i1})\varpi_1 = R_{i1}\varpi_1 + \tau_i \varpi_1$

The calculation $S_2$ for the session key $E(h_2, h_{id})$ yields the same result. The calculation of the submatrices $\varpi_j$ is $j=\overline{2,q-1}$ determined by the parameterization of the arrays $\beta_{ij}$, $i=\overline{1,l}$, $j=\overline{1,q}$ and expression (21) for $S_1$ and $S_2$, respectively.

$$\varpi_{j1} = \gamma_j \varpi_q(h_1) \lambda_j^{-1} + \delta_j \varpi_1 \lambda_j^{-1}$$
$$\varpi_{j2} = \gamma_j \varpi_q(h_2) \lambda_j^{-1} + \delta_j \varpi_1 \lambda_j^{-1} \quad , \quad (22)$$

where $j=\overline{2,q-1}$. Expression (22) is linear with respect to the parameter $\varpi_q$. The parameters $\gamma_j, \delta_j, \lambda_j, \varpi_1$ are secret and fixed. For $\varpi_q(h_1) \neq \varpi_q(h_2)$ we obtain $\varpi_{j1} \neq \varpi_{j2}$, $E(h_1, h_{id}) \neq E(h_2, h_{id})$, and the shared secret $S$ is unique across sets of hashes $h_1$, $h_{id}$ and $h_2$, $h_{id}$.

## 5      Security Analysis

Let's consider brute force attacks. We consider *signature forgery attack.*

The attack is aimed at modifying the message *msg'* and is based on the selection of session keys $\psi(h_1)$ and $\psi(h_2)$. Two attack vectors are possible.

***The first attack*** is based on enumeration of values $r_1'$, $r_2'$ such that a collision is found $\psi_q(h_1') = \psi_q(h_1)$ and $\psi_q(h_2') = \psi_q(h_2)$, where $h_1' = H(r_1', msg')$, $h_2' = H(r_2', msg')$. This will allow session keys to be used $\psi(h_1)$ to $\psi(h_2)$ forge a signature, since it is possible to calculate $x' = S^{-1}(y)$, where, based on the shared secret $y = H(msg')$. The tuple of signature parameters for *msg'* will be equal to $sig = (x', \psi_1, \psi_2, r_1', r_2')$. The probability of success of the attack is equal to

$$P_\psi = 2^{-2m^2}.$$

If you use signature with $t$ session keys, the probability of success will be

$$P_{\psi,t}=2^{-tm^2}.$$

The secrecy against signature forgery attack will be equal to
$$C_{\psi,t}=tm^2.$$

**The second attack** is based on enumerating the values of $r_1'$, $r_2'$, calculating $h_1'=H(r_1',msg')$, $h_2'=H(r_2',msg')$ and constructing session keys $\psi(h_1')$ and $\psi(h_2')$. For $\psi(h_1')$ and $\psi(h_2')$ the shared secret is calculated. This shared secret is a two permutation vectors $S_1=B\cdot\psi(h_1')$, $S_2=B\cdot\psi(h_2')$ and two input vectors $x_1=S_1^{-1}(y)$, $x_2=S_2^{-1}(y)$, where $y=H(msg')$. Verification $x_1=x_2$ is equivalent to checking the condition $y_1=y_2=H(msg')$. The signature for $msg'$ will have a tuple of parameters $sig=(x',\psi(h_1'),\psi(h_2'),r_1',r_2')$. The length of the vectors $x_1,x_2$ is $L_x=ml$. The probability that two vectors will coincide with randomly selected ones $\psi(h_1')$ will $\psi(h_2')$ be equal to
$$P_x=2^{-ml}.$$

If you use signature with $t$ session keys, the probability of success will be
$$P_{x,t}=2^{-ml(t-1)}.$$

The secrecy against signature forgery attack will be equal to
$$C_{x,t}=ml(t-1).$$

The condition check $x_1=x_2$ can be replaced by a check for equality of secrets $S_1=S_2$. The size of the shared secret $S$ is equal to $L_S=2m^2l$. The success probability of a forgery attack through matrix comparison $S_1$ will $S_2$ be equal to
$$P_S=2^{-2m^2l}.$$

If you use signature with $t$ session keys, the probability of success via shared secret verification will be
$$P_{S,t}=2^{-2m^2l(t-1)},$$

c secrecy will be equal $C_{S,t}=2m^2l(t-1)$

The bit size $t$ of session keys $\psi$ is $L_{\psi,t}=tm^2q$.

The reliability of identity verification through equality of secrets $S$ with generated random session keys is determined by the condition
$$L_S \geq L_{\psi,t}$$
$$2m^2l \geq tm^2q$$
$$l \geq tq/2$$

The probability of successfully guessing the session key $\psi$ for one attempt will have a probability

$$P_\psi = 2^{-m^2(q-1)}.$$

The expression for $P_\psi$ takes into account that in the session key $\psi$ it is necessary to guess only $q-1$ the sub-matrices, since the last sub-matrix is not secret.

To prove identity, you need to guess $t$ the session key, and the secrecy in this case will be equal to $C_{\psi,t} = tm^2(q-1)$.

***Collision attack.*** The possibility of a collision attack is determined by the fact that session signature keys $\psi(h)$ are built by hashes $h = H(r, msg)$ The probability of success of a collision attack depends on the accumulated key base $N$, the size of the submatrices $\varpi_q(h)$, and the number of session keys $t$.

$$P_{col} = N * 2^{-tm^2}$$

The secrecy against collision attack will be equal to

$$C_{col} = tm^2 - \log N \text{ bit.}$$

If the signer creates and transmits to the verifier one signature per second, then over 100 years the size of the signature database $N = 2^{32}$ and

$$C_{col} = tm^2 - 32.$$

An analytical attack on the LINEture algorithm is aimed at factorizing the matrix $\psi$ to obtain a long-run matrix $\omega$, the substitution vector $\beta$, the parameters $\gamma_j, \delta_j, \lambda_j$ and the matrix $\varpi_1$ that is generated in each signature.

***An analytical attack on session key factorization.*** The attack algorithm consists of the following steps. The public session key is determined by expression (9)

$$\psi[mq \times m] = \omega^{-1} E^T.$$

Factorization $\psi$ into matrices $\omega$ will $E$ allow us to construct a permutation vector

$$\beta = B\omega^{-1}.$$

Let us consider factorization $\psi$ into matrices $\omega$ and $E^T$.

Let us consider the case when $m=2$ and $q=2$. The matrix $\psi$ has dimension $[4 \times 4]$, $E^T$ - dimension $[4 \times 2]$. Let us write the system of equations for the product $\omega^{-1}\tilde{E}^T$

$$\psi = \omega^{-1} \times E^T = \omega^{-1} \times \begin{vmatrix} \varpi_1 \\ \varpi_2 \end{vmatrix} = \begin{vmatrix} \omega_{11} & \omega_{12} & \omega_{13} & \omega_{14} \\ \omega_{21} & \cdot & \cdot & \cdot \\ \omega_{31} & \cdot & \cdot & \cdot \\ \omega_{41} & \cdot & \cdot & \omega_{44} \end{vmatrix} \times \begin{vmatrix} \varpi_{11} & \varpi_{12} \\ \varpi_{21} & \varpi_{22} \\ \varpi_{31} & \varpi_{32} \\ \varpi_{41} & \varpi_{42} \end{vmatrix} =$$

$$\begin{vmatrix} \omega_{11}\varpi_{11} + \omega_{12}\varpi_{21} + \omega_{13}\varpi_{31} + \omega_{14}\varpi_{41} & \omega_{11}\varpi_{12} + \omega_{12}\varpi_{22} + \omega_{13}\varpi_{32} + \omega_{14}\varpi_{42} \\ \omega_{21}\varpi_{11} + \omega_{22}\varpi_{21} + \omega_{23}\varpi_{31} + \omega_{24}\varpi_{41} & \omega_{21}\varpi_{12} + \omega_{22}\varpi_{22} + \omega_{23}\varpi_{32} + \omega_{24}\varpi_{42} \\ \omega_{31}\varpi_{11} + \omega_{32}\varpi_{21} + \omega_{33}\varpi_{31} + \omega_{34}\varpi_{41} & \omega_{31}\varpi_{12} + \omega_{32}\varpi_{22} + \omega_{33}\varpi_{32} + \omega_{34}\varpi_{42} \\ \omega_{41}\varpi_{11} + \omega_{42}\varpi_{21} + \omega_{43}\varpi_{31} + \omega_{44}\varpi_{41} & \omega_{41}\varpi_{12} + \omega_{42}\varpi_{22} + \omega_{43}\varpi_{32} + \omega_{44}\varpi_{42} \end{vmatrix}$$

The number of unknowns of the matrix $\omega$ is equal to $m^2q^2=16$, and for the matrix $E^T$ $m^2q=8$ The total number of unknowns is 24. The number of equations $m^2q=8$ and the uncertainty regarding the solutions for binary matrices will be $m^2q^2=16$. The number of equations can be increased by $m^2q=8$ using the equations for the second session key $\psi_2$. Session keys $\psi_1$ and $\psi_2$ have identical matrices $\varpi_1$ and different ones $\varpi_2$. Adding equations using the session key $\psi_2$ will increase $m^2(q-1)=4$ the number of unknowns due to the matrix $\varpi_2$, but due to the identical parts in the equations for the unknown coefficients, $\varpi_1$ the rank of the matrix will decrease by $m^2=4$. For 28 unknowns and 16 equations, we obtain an uncertainty regarding the solutions for binary matrices equal to $m^2(q^2-1)=12$. Adding $2m^2q=16$ equations using session keys for other signatures does not reduce the uncertainty regarding the solutions, since the rank of the matrix of the system of equations decreases by $m^2=4$. The uncertainty regarding the solutions for binary matrices $\omega$ and $E$ will always be equal to $m^2(q^2-1)$. It can be concluded that factorization $\psi$ into matrices $\omega$ and $E$ has only brute-force complexity, which has the following complexity estimates.

The matrix $\omega$ constructed using the transformation 2 has dimension $[m(q-1)\times m(q-1)]$. An attack through solving a system of quadratic equations $\psi=\omega^{-1}E^T$ considering the matrix dimension $\omega$ has $E^T$ complexity

$$N_{\omega,E}=2^{m^2(q-1)^2-m^2}=2^{m^2(q^2-2q)}.$$

Brute-force attack on factorization $\psi$ has complexity

$$N'_{\omega,E}=2^{m^2(q-1)^2+m^2(q-1)}=2^{m^2(q^2-q)}.$$

The secrecy against an attack through solving systems and equations and a brute-force attack will be equal to

$$C_{\omega,E}=m^2(q-1)^2-m^2=m^2(q^2-2q) \text{ And}$$
$$C'_{\omega,E}=m^2(q-1)^2+m^2(q-1)=m^2(q^2-q)$$

For the parameters $m=8$, $q=3$ we obtain the secrecy $C_{\omega,E}=m^2(q^2-2q)=64*3=192$ bit and $C'_{\omega,E}=m^2(q^2-q)=64*6=384$ bit.

The next step is to perform parametric factorization $\beta$ on $\tau, \gamma_j, \delta_j, \lambda_j$, $j=\overline{2,q-1}$. The substitutions $\beta_{ij}$, $i=\overline{1,l}$, $j=\overline{2,q-1}$ are defined by expressions (12), (13)

$$\beta_{i1}=R_{i1}+\tau_i+\sum_{j=2}^{q-1}R_{ij}\delta_j,$$
$$\beta_{ij}=R_{ij}\gamma_j$$

$$\beta_{iq} = \sum_{j=2}^{q-1} R_{ij} \lambda_j$$

where $R_{ij}[2m \times m]$, $j=\overline{2,q-1}$ are binary matrices, are random and secret. The matrix $R_{i1}[2m \times m]$ is secret, constructed using the formulas of transformation 1. Factorization of permutations $\beta_{ij}$ has no solution due to the randomness of the factors in the products.

Let's consider the cost estimates for implementation, digital signature, identity proof, and performance of the LINEture cryptosystem.

*Implementation costs* are determined by the costs of general parameters, public and private keys.

**General parameters of the cryptosystem:** $m,l,q,t$ - 4 bytes

**Private signature key** sk : $\omega$ , $\beta$

- matrix $\omega[mq \times mq]$ contains $\omega_1[m(q-1) \times m(q-1)]$, constructed by a generator based on the initial parameter, with a total size $n_\omega=32$ byte:

- the vector $\beta=[\beta_1,\beta_2,...,\beta_l]$ includes $l$ submatrices $\beta_i[2m \times mq]=\beta_{i1}\|\beta_{i2}\|...\|\beta_{iq}$ with a total size $n_\beta=2lm^2q/8$ of bytes.

- matrices $\delta_j[m \times m]$, $\lambda_j[m \times m]$, $\gamma_j[m \times m]$, $j=\overline{2,q-1}$ are constructed by the generator based on the initial parameter, with a total size $n_\delta=32$ of a byte

- matrices $\tau=[\tau_1,\tau_2,...,\tau_l]$, $\tau_i[m \times m]$, $i=\overline{1,l}$ with a total size $n_\tau=lm^2/8$ of bytes.

**Public signature key** pk **:** $B$

- the vector $B$ includes $l$ matrices $\beta_i[2m \times mq]$ with a total size $n_B=2lqm^2/8$ of bytes

**Signature** $sig=(x,\psi_v,r_v), v=\overline{2,t}$

- the vector $x$ includes $l$ words with a total size $n_x=lm/8$ of bytes.

- each vector $\psi_v$ includes $v=\overline{2,t}$ submatrices $q$ with $\psi_i[m \times m]$ a total size $n_\psi=tqm^2/8$ of bytes.

- $r_v$, $v=\overline{2,t}$ bit strings of 32 bytes, with a total size $n_r=32t$ of a byte

Table 3 presents the estimated costs in bytes for common parameters, keys, and digital signature.

**Table 3.** Cost for general parameters, keys and digital signature.

| General parameters: $m,l,q,t$ (4 bytes) | Private key: $\omega$ (32 bytes), $\tau, \gamma_j, \delta_j, \lambda_j$ (32 bytes), $\beta$ ( $n_\beta=2lm^2q/8$ byte) | | Public key: $B$ ( $n_B=2lqm^2/8$ byte) | Signature: $sig=(x,\psi_v,r_v), v=\overline{2,t}$ $n_x=lm/8$ $n_\psi=tqm^2/8$ $r_v$ - $n_r=32t$ byte | | | |
|---|---|---|---|---|---|---|---|
| $m,l,q,t$ | $\beta$ | $\Sigma$ | $B$ | $x$ | $n_\psi$ | $n_r$ | $\Sigma$ |
| 8.8, 4,2 | 512 | 576 | 512 | 8 | 32 | | 160 |

| | | | | | | |
|---|---|---|---|---|---|---|
| 8.8, 3, 2 | 384 | 448 | 384 | 8 | 48 | 64 | 120 |
| 8,8,3,3 | 384 | 448 | 384 | 8 | 72 | 96 | 176 |
| 8.8, 3, 4 | 384 | 448 | 384 | 8 | 96 | 128 | 232 |
| 8.8, 3, 5 | 384 | 448 | 384 | 8 | 120 | 160 | 288 |
| 8, 16, 3, 2 | 768 | 832 | 768 | 16 | 48 | 64 | 128 |
| 8, 16, 3, 3 | 768 | 832 | 768 | 16 | 72 | 96 | 184 |
| 8, 16, 3, 4 | 768 | 832 | 768 | 16 | 96 | 128 | 240 |
| 16,8, 3,2 | 1536 | 1600 | 1536 | 16 | 192 | 64 | 272 |
| 16,8, 3,3 | 1536 | 1600 | 1536 | 16 | 288 | 96 | 400 |

Table 4 presents the secrecy estimates in bits.

**Table 4.** Cost for general parameters, keys and digital signature.

| General parameters | Key and signature parameters | Analytical attack $C_{\omega,E}=m^2(q^2-2q)$ $C'_{\omega,E}=m^2(q^2-q)$ | | Signature forgery attack $C_{\psi,t}=tm^2$ $C_{x,t}=ml(t-1)$ $C_{S,t}=2m^2l(t-1)$ | | | Collision attack $C_{col}=tm^2-32$ |
|---|---|---|---|---|---|---|---|
| $m, l, q, t$ $l \geq tq/2$ 4,5,6,8 | Welcome \ Open \ Sub | $C_{\omega,E}$ | Selective attack $C'_{\omega,E}$ | $C_{\psi,t}$ | $C_{x,t}$ | $C_{S,t}$ | $C_{col}$ |
| 8.8,3,2 | 448\384\120 | 192 | 384 | 128 | 64 | 1024 | 96 |
| 8,8,3,3 | 448\384\176 | 192 | 384 | 192 | 128 | 2048 | 160 |
| 8,8,3,4 | 448\384\232 | 192 | 384 | 256 | 192 | 3072 | 224 |
| 8,8,3,5 | 448\384\288 | 192 | 384 | 320 | 256 | 4095 | 288 |
| 8,16,3,2 | 832\768\128 | 192 | 384 | 128 | 128 | 2048 | 96 |
| 8,16,3,3 | 832\768\184 | 192 | 384 | 192 | 256 | 4096 | 160 |
| 8,16,3,4 | 832\768\240 | 192 | 384 | 256 | 384 | 6144 | 224 |
| 16,8,3,2 | 1600\1536\272 | 768 | 1536 | 512 | 128 | 4096 | 480 |
| 16,8,3,3 | 1600\1536\400 | 768 | 1536 | 768 | 256 | 4096 | 736 |

A collision attack has the least security. Practical implementation relies on the ability to accumulate and use a database of hashes and session keys. The effect of accumulated hashes is limited due to the practical impossibility of accumulating a significantly larger database. Signature forgery attack secrecy for 8-bit substitutions has 128\192\256 bits of secrecy. Table 5 shows comparative characteristics of the costs of keys and signatures in bytes with known cryptosystems [32].

**Table 5.** Comparison of key and signature costs in bytes.

| Version | Security Level | Private Key Size | Public Key Size | Signature Size |
|---|---|---|---|---|
| ML-DSA-44 | AES128 | 2560 | 1312 | 2420 |
| ML-DSA-65 | AES192 | 4032 | 1952 | 3309 |
| ML-DSA-87 | AES256 | 4896 | 2592 | 4627 |
| RSA3072 | AES128 | 384 | 384 | 384 |
| RSA-7680 | AES192 | 960 | 960 | 960 |
| RSA15360 | AES256 | 1920 | 1920 | 1920 |
| LINEture 128<br>m=8, l= 86, q =3, t=3 | AES128 | 448 | 384 | 176 |
| LINEture 192<br>m=8, l= 8 6, q =3 , t=4 | AES192 | 448 | 384 | 232 |
| LINEture 256<br>m=8, l= 8 6, q =3, t=5 | AES 256 | 448 | 384 | 288 |

## 6      Conclusions

The LINEture cryptosystem exploits the concept of a brute-force search problem. The cryptosystem's security is determined by the fact that the matrices of the public key, session key, and shared secret signature have a rank lower than that of the secret matrix of the private key, rendering the inverse problem of private key recovery infeasible.

The LINEture cryptosystem is a strong candidate for post-quantum cryptography. By selecting the general parameters of the cryptosystem, it can implement the NIST-declared security levels of 128, 192, and 256 bits, as well as any other levels. The LINEture algorithm scales well across computational costs, memory, and hardware platform limitations without sacrificing high security. The sizes of private and public keys for computations on 8-bit words are 448 and 384 bytes, respectively, while the signature size is 176, 232, or 288 bytes, depending on the security level. This demonstrates superior results compared to other post-quantum candidates. A software implementation of the LINEture algorithm using vector-based bitwise matrix computations can be highly efficient.